\pdfoutput=1

\documentclass[aps,twocolumn,preprintnumbers,amssymb,nobibnotes,nofootinbib,longbibliography,superscriptaddress,10pt]{revtex4-2}

\usepackage[utf8]{inputenc}

\usepackage{graphicx,epsf}
\usepackage{amsmath,amsfonts,amssymb,amsbsy,mathrsfs}
\usepackage{mathtools}
\usepackage{slashed}
\usepackage{comment}
\usepackage{color}
\usepackage{subcaption}
\usepackage{hyperref}[colorlinks, allcolors=black]
\usepackage[capitalise]{cleveref}

\makeatletter
\g@addto@macro\bfseries{\boldmath}
\makeatother

\newcommand{\eps}{\epsilon}

\newcommand{\ii}{\mathrm{i}}
\newcommand{\dd}{{\mathrm{d}}}

\begin{document}

\title{Canonical Differential Equations for the Elliptic Two-Loop Five-Point Integral Family Relevant to $t\bar t +$Jet Production at Leading Color}

\preprint{DESY 25-029, ZU-TH 13/25}

\author{Matteo~Becchetti}
\affiliation{Dipartimento di Fisica e Astronomia, Università di Bologna e INFN, Sezione di Bologna, via Irnerio 46,
I-40126 Bologna, Italy}
\author{Christoph~Dlapa}
\affiliation{Deutsches Elektronen-Synchrotron DESY, Notkestr.\ 85, 22607 Hamburg, Germany}
\author{Simone~Zoia}
\affiliation{Physik-Institut, University of Zurich, Winterthurerstrasse 190, 8057 Zurich, Switzerland}

\begin{abstract}

We complete the construction of canonical differential equations (DEs) for all families of Feynman integrals appearing in the two-loop leading-color QCD amplitude for top-pair production in association with a jet at hadron colliders.
This allows us to obtain analytic results for all required Feynman integrals in terms of iterated integrals with closed-form kernels.
To achieve this, we study the two-loop five-point integral family for which canonical DEs were previously unavailable due to the appearance of elliptic functions and nested square roots.
In addition to marking a significant step towards next-to-next-to-leading-order QCD predictions for a high-priority LHC process, this is the first time that canonical DEs are obtained for Feynman integrals where elliptic functions appear in conjunction with the high algebraic complexity of a process with more than four particles.
Our results also reveal a number of new interesting analytic features, such as a `duplet' structure that generalizes the even/odd parity of square roots to nested square~roots.

\end{abstract}

\maketitle

Feynman integrals are central in the study of quantum field theory (QFT).
Not only are they essential to obtain precise predictions for particle phenomenology,
but they also drive formal studies, and offer increasingly many links to mathematics and other areas of physics.
Progress in their computation is crucial for developments across these fields.
However, this progress is challenged by both algebraic and analytic complexity.
The former grows rapidly with the number of variables and thus of scattered particles, whereas the latter escalates as new classes of special functions emerge.
In this Letter, we present a computation that lies at the cutting edge of both types of complexity by obtaining---for the first time---analytic results for a family of two-loop five-particle Feynman integrals involving elliptic functions and nested square roots.

One of the most powerful frameworks for computing Feynman integrals consists in viewing them as the solution to certain differential equations (DEs)~\cite{Barucchi:1973zm,Kotikov:1990kg,Kotikov:1991hm,Gehrmann:1999as,Bern:1993kr,Henn:2013pwa}.
The insight that a \emph{canonical} form of the DEs~\cite{Henn:2013pwa}---one which dramatically simplifies the solution---exists has made possible increasingly many results in high-energy physics.
In the best understood case, canonical means that the dependence on the regulator of dimensional regularization is factorized, and the DEs contain solely logarithmic differential one-forms ($\dd \log$'s).
In such a case, we have a well developed mathematical technology to handle the solution (see Refs.~\cite{Henn:2014qga,Abreu:2022mfk,Weinzierl:2022eaz,Badger:2023eqz} for reviews), and the state of the art has reached integrals with five~\cite{Liu:2024ont} and six particles~\cite{Henn:2024ngj,Abreu:2024fei,Henn:2025xrc} at three- and two-loop order, respectively.

However, starting at two-loop order, Feynman integrals may involve transcendental functions linked to higher-genus geometries~\cite{Sabry:1962rge,Bourjaily:2022bwx}, such as periods of elliptic curves, whose DEs cannot be expressed in the form above.
Such functions are typically linked to the presence of massive virtual particles.
Despite recent progress in extending the construction of canonical DEs~\cite{Adams:2018yfj,Frellesvig:2021hkr,Gorges:2023zgv,Dlapa:2022wdu,Pogel:2022ken,Pogel:2022vat,Gorges:2023zgv,Driesse:2024feo,Duhr:2024uid},
the most advanced applications are confined to processes with comparatively simpler four-particle kinematics~\cite{Duhr:2021fhk,Muller:2022gec,Gorges:2023zgv,Delto:2023kqv,Duhr:2024bzt,Schwanemann:2024kbg,Becchetti:2025rrz}.
Advancing our methodology to handle high-multiplicity processes with massive virtual particles is imperative.
This is one of the main bottlenecks to keep the theoretical uncertainties in line with the experimental ones for many processes in the Large Hadron Collider (LHC) program~\cite{Andersen:2024czj}.
Moreover, thanks to their rich kinematics, high-multiplicity processes offer an extraordinary laboratory for uncovering new properties of QFT.

We break through the wall of elliptic integrals with five-point kinematics, and obtain canonical DEs for a family of Feynman integrals relevant for top-pair production with a jet at hadron colliders that involves elliptic functions~\cite{Badger:2022hno,Badger:2024fgb,Badger:2024dxo}.
Our result completes the construction of canonical DEs for all families appearing in the leading-color two-loop QCD amplitude for this process, allowing us to express all required Feynman integrals analytically in terms of iterated integrals~\cite{Chen:1977oja}, and thus marking a key step towards obtaining next-to-next-to-leading-order (NNLO) QCD predictions.
This achievement is of immediate importance for the LHC, where top-pair production with a jet is a high-priority process~\cite{CMS:2016oae,ATLAS:2018acq,CMS:2020grm,CMS:2024ybg}.
Its high sensitivity to the top-quark mass~\cite{Alioli:2013mxa,Bevilacqua:2017ipv,Alioli:2022ttk,Alioli:2022lqo} and the increasing experimental precision call for the NNLO QCD corrections to be computed.
Our results advance the theoretical understanding of elliptic Feynman integrals to a new, high-multiplicity regime, provide analytical data to fuel ongoing formal studies, and open new research avenues.

\section{Setting the stage} \label{sec:stage}

\begin{figure}[t!]
		\includegraphics[width=0.4\linewidth]{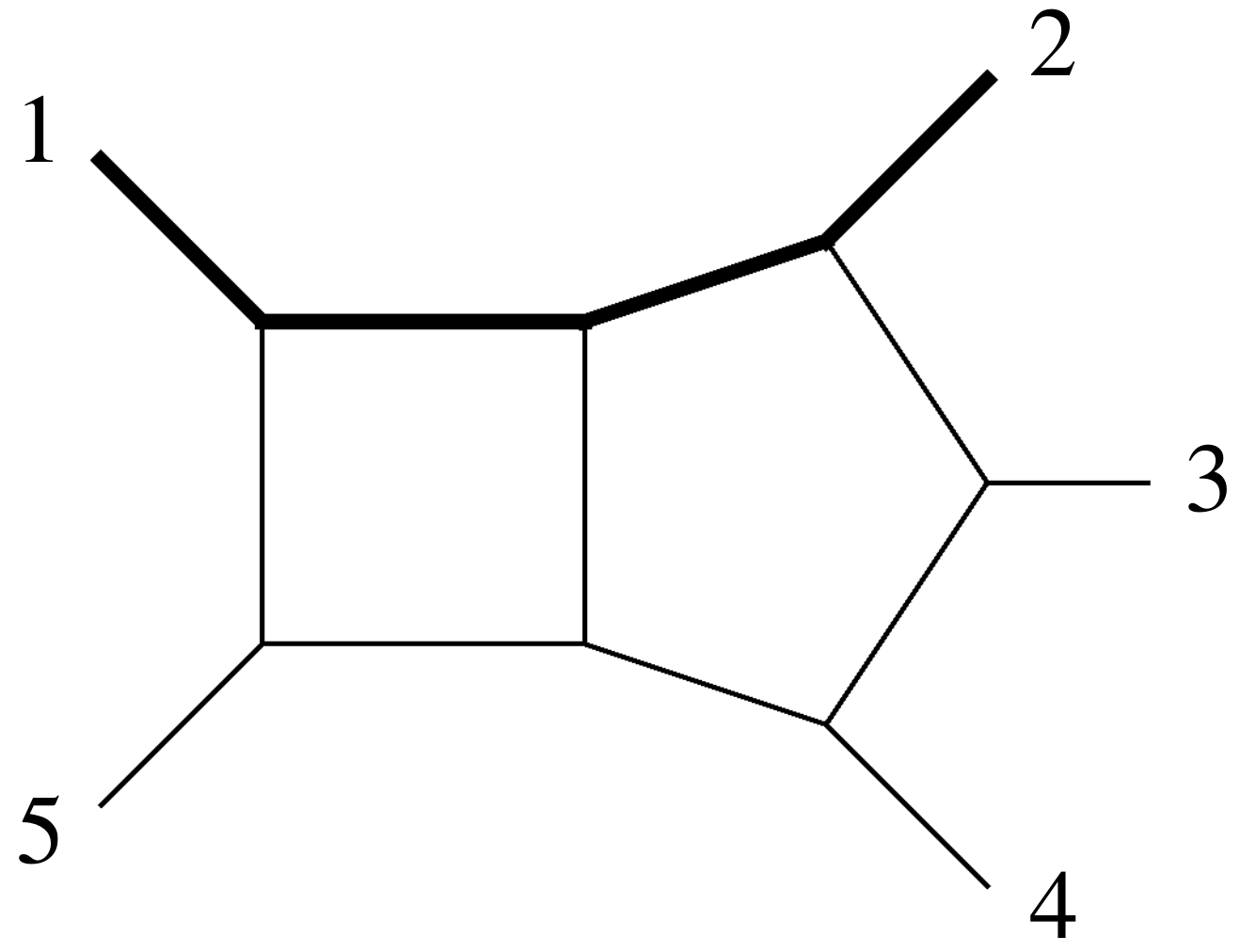}
		\label{fig:PBttjB}
 \vspace{0.2cm}
        \caption{Graph of the integral family studied in this Letter. Thin (thick) lines are massless (massive).}
	\label{fig:topo}
\end{figure}

We consider the `pentagon-box' integral family shown in \cref{fig:topo}.
We use the notation of Ref.~\cite{Badger:2024fgb}:
the external momenta $p_i$ are outgoing and on shell, i.e., $p_1^2 = p_2^2 = m_t^2$ and $p_3^2 = p_4^2 = p_5^2=0$, and the kinematics are described by six independent Lorentz invariants,
\begin{align}
\label{eq:x}
\vec{x}=\bigl( d_{12}, d_{23}, d_{34}, d_{45}, d_{15}, m_t^2 \bigr) \,,
\end{align}
where $d_{ij} = p_i \cdot p_j$.
The family is the set of all scalar integrals of the form
\begin{equation}
  I_{a_1,\ldots,a_8}^{a_9,a_{10},a_{11}}(\vec{x},\eps) = \mathrm{e}^{2\eps \gamma_{\rm E}} \int \frac{\dd^d k_1}{\ii \pi^\frac{d}{2}}  \frac{\dd^d k_2}{\ii \pi^\frac{d}{2}} 
  \frac{D_{9}^{a_9} D_{10}^{a_{10}} D_{11}^{a_{11}}}{D_{1}^{a_1}\cdots D_{8}^{a_8}}\,,
  \label{eq:topodef}
\end{equation}
where $a_9,a_{10},a_{11} \geq 0$, $d=4-2\eps$, and the inverse propagators $D_i$ are given in Table~1 of Ref.~\cite{Badger:2024fgb} (family ${\rm PB}_B$).
Modulo integration-by-parts~\cite{Tkachov:1981wb,Chetyrkin:1981qh,Laporta:2000dsw} and symmetry relations, the family has a basis of 121 master integrals (MIs).
The analytic structure of the integrals involves six square roots: $\sqrt{\beta_{12}}$, $\sqrt{\Delta_{31}}$, $\sqrt{\Delta_{32}}$, $\sqrt{\Delta_{33}}$, $\sqrt{\Lambda_4}$, $\sqrt{\Delta_5}$.\footnote{With respect to Ref.~\cite{Badger:2024fgb}, we make the square-root sign explicit, and use $\sqrt{\Delta_5}$ rather than $\mathrm{tr}_5$ to avoid confusion with the homonymous pseudo-scalar invariant.}

\begin{figure}
	\centering
	\begin{subfigure}{0.30\linewidth}
		\includegraphics[width=\columnwidth]{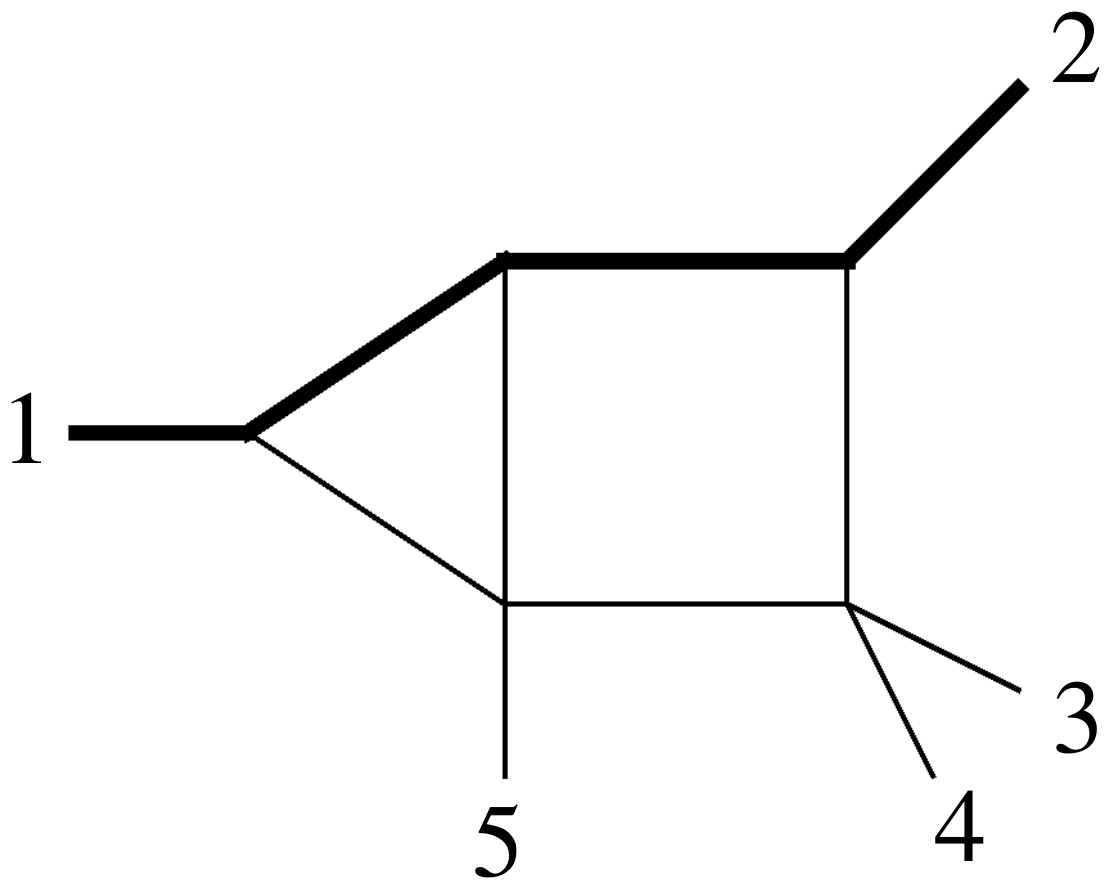}
		\caption{MIs \#35--37.}
		\label{fig:elliptic}
	\end{subfigure}
	\hspace{1cm}
	\begin{subfigure}{0.30\linewidth}
		\includegraphics[width=\columnwidth]{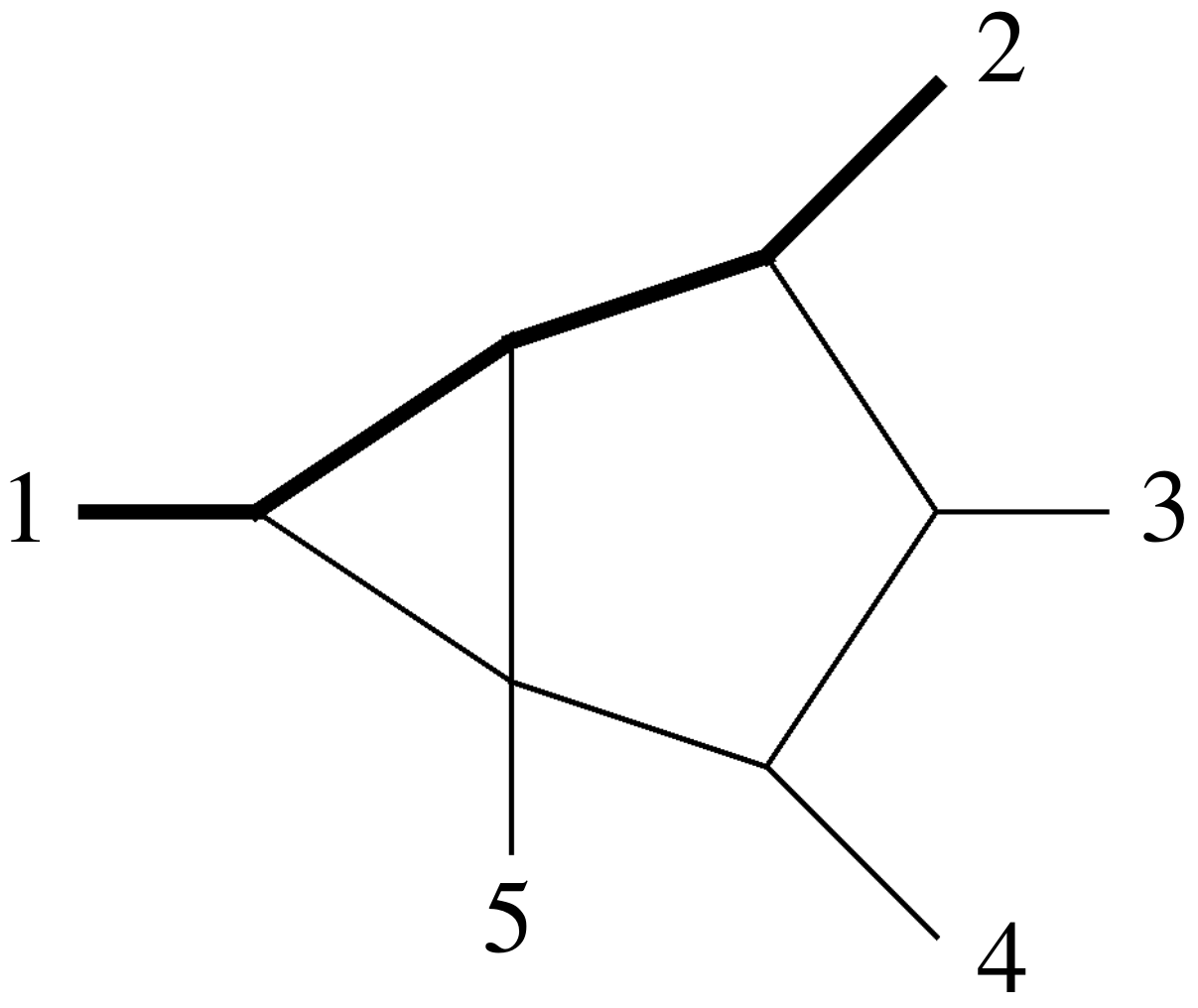}
		\caption{MIs \#18--20.}
		\label{fig:nested}
	\end{subfigure}
	\caption{Sectors whose maximal cuts introduce the elliptic curve (a), and the nested square roots (b), and labels of the corresponding MIs.}
	\label{fig:probSec}
\end{figure}

In general, a set of MIs $\vec{\mathcal{I}}$ satisfies a system of differential equations (DEs) of the form~\cite{Barucchi:1973zm,Kotikov:1990kg,Kotikov:1991hm,Gehrmann:1999as,Bern:1993kr}
\begin{align}
\label{eq:DEsGeneral}
\dd \, \vec{\mathcal{I}}(\vec{x}, \eps) = \dd A\left(\vec{x},\eps\right) \cdot \vec{\mathcal{I}}(\vec{x}, \eps)\,,
\end{align}
where $\dd$ is the total differential with respect to $\vec{x}$.
The connection matrix $\dd A(\vec{x},\eps)$ for the basis $\vec{\mathcal{I}}$ of Ref.~\cite{Badger:2024fgb} depends quadratically on $\eps$, and contains non-logarithmic one-forms.
These non-canonical features are however localized in the two sectors shown in \cref{fig:probSec}, i.e., in the subsets of integrals having the propagators of the graphs in the figure.
More concretely, the `homogeneous' blocks---i.e., the blocks on the diagonal of the connection matrix that couple the MIs of the same sector---were cast into `$\eps \times \dd \log$' form for all sectors but the two in \cref{fig:probSec}.

As noticed in Ref.~\cite{Badger:2024fgb}, the homogeneous block of the sector in \cref{fig:nested} is $\eps$-factorized by rotating $\mathcal{I}_{19}$ and $\mathcal{I}_{20}$~as
\begin{align} \label{eq:nested_transf}
 \mathcal{J}_{19} = \frac{\sqrt{n_+}}{d_{45} \, r_2} \bigl(\mathcal{I}_{20} + \mathcal{I}_{19}   \bigr) \,, \
 \mathcal{J}_{20} = \frac{\sqrt{n_-}}{d_{45} \, r_2} \bigl(\mathcal{I}_{20} - \mathcal{I}_{19}   \bigr)\,.
\end{align}
This transformation contains the nested square roots
\begin{align} \label{eq:npm}
\sqrt{n_{\pm}} = \sqrt{d_{23}^2 \Delta_5 - 8 r_2 r_4 \pm 4 d_{23} r_3 \sqrt{\Delta_5}} \,,
\end{align}
where $r_k$ denotes a degree-$k$ polynomial in $\vec{x}$.
The resulting $\eps$-factorized block was not written in terms of $\dd \log$'s, and it is unclear whether this is generally possible in the presence of nested roots~\cite{FebresCordero:2023pww}.
We will show that this is indeed possible in our case, and discuss the symmetries associated with the nested square roots.
Note that the interior square root, $\sqrt{\Delta_5}$ in \cref{eq:npm}, can be rationalised, e.g., by using momentum-twistor variables~\cite{Badger:2022mrb}, at the cost of increasing the algebraic complexity of the expressions and obscuring the symmetry properties discussed in \cref{sec:AnalyticStructure}.
We therefore refrain from doing such a change of variables.

The sector in \cref{fig:elliptic} was shown to involve elliptic functions associated with the elliptic curve $y^2 = \mathcal{P}(z)$,
\begin{align}
\mathcal{P}(z) = (z-e_1) (z-e_2) (z-e_3) (z-e_4) \,,
\end{align}
where the roots $e_i$ are functions of $\vec{x}$.
In particular, the maximal cut of the scalar integral of this sector (${\cal I}_{35}$, with a suitable normalization) at $\eps=0$ is given, depending on the integration cycle, by the periods of the elliptic curve~\cite{Badger:2024fgb}.
We choose them as
\begin{equation}\label{eq:periods}
 \begin{aligned}
  \psi_1=\frac{2}{\pi}\int_{e_2}^{e_3}\frac{\dd z}{\sqrt{\mathcal{P}(z)}}&=\frac{4 \,\mathrm{K}(\kappa^2)}{\pi\sqrt{(e_3-e_1)(e_4-e_2)}} \,,\\
  \psi_2=4\ii \int_{e_1}^{e_2}\frac{\dd z}{\sqrt{\mathcal{P}(z)}}&=\frac{-8 \, \mathrm{K}(1-\kappa^2)}{\sqrt{(e_3-e_1)(e_4-e_2)}} \,,
 \end{aligned}
\end{equation}
where $\mathrm{K}$ is the complete elliptic integral of the first kind,
\begin{equation} \label{eq:K}
 \mathrm{K}(\kappa^2)=\int_0^1\frac{\dd t}{\sqrt{(1-t^2)(1-\kappa^2t^2)}} \,,
\end{equation}
$\kappa^2 = e_{32} e_{41}/(e_{31} e_{42})$, with $e_{ij} = e_i-e_j$, is the elliptic modulus, and the normalizations in \cref{eq:periods} were chosen to simplify the $m_t^2$ expansion.
The presence of the elliptic curve is the main obstacle towards achieving canonical DEs, which we overcome in the next section.

\section{Construction of a Canonical Basis}
\label{sec:CanonicalConstruction}

There is no consensus yet on what
`canonical' means in addition to the factorization of $\eps$ beyond the $\dd \log$ case~\cite{Broedel:2018qkq,Frellesvig:2023iwr,Duhr:2024uid}. 
We adopt a \emph{local} definition: we require that, for each singular point, it is possible to choose the appearing period of the elliptic curve such that all one-forms have at most simple poles.  
In this section, we will be concerned with achieving the factorization of $\eps$.
We will verify that the resulting DEs fulfil this notion of canonical in the next Section.
We denote by $T(\vec{x},\eps)$ the transformation matrix from the basis $\vec{\mathcal{I}}$ of Ref.~\cite{Badger:2024fgb} to the new, canonical basis $\vec{\mathcal{J}}$, i.e., $\vec{\mathcal{J}}=T\cdot \vec{\mathcal{I}}$.
We include in $T$ the transformation in \cref{eq:nested_transf}.

We construct $T$ in two steps. 
First, we transform the homogeneous block of the DEs related to the elliptic sector into $\eps$-factorized form.
Second, we perform a similar rotation to factorize $\eps$ in the inhomogeneous blocks that couple other sectors to the ones shown in \cref{fig:probSec}.
In both cases, we follow the approach of Ref.~\cite{Gorges:2023zgv}.
An important part of this method is to find a suitable starting basis that simplifies the DEs as much as is possible without introducing transcendental functions. 
We find the basis of Ref.~\cite{Badger:2024fgb} to be a suitable starting point. In particular, the maximal cut of $\mathcal{I}_{35}$ is related to the periods of the elliptic curve (see Appendix~A of Ref.~\cite{Badger:2024fgb} and \cref{eq:period-cut}). 

We begin by focusing on the derivative w.r.t.~$m_t^2$. 
The homogeneous block of the connection matrix of the elliptic sector takes the following form at $\eps=0$,
\begin{equation}\label{eq:elliptic-ep0}
 \dd A^{\eps=0}_{\text{hom.,ell.}}= A^{\eps=0}_{m_t^2,\text{hom.,ell.}} \, \dd m_t^2 + \ldots \,,
\end{equation}
where we omit the differentials in the other variables, and
\begin{equation} \label{eq:elliptic-ep0bis}
 A^{\eps=0}_{m_t^2,\text{hom.,ell.}}=\begin{pmatrix}
                                                    B & \begin{matrix}
                                                    0 \\ 0 \end{matrix}\\
                                                    \begin{matrix}
                                                    \star & \star
                                                    \end{matrix} & 0
                                                   \end{pmatrix} \,,
\end{equation}
with a $2\times 2$ block $B$ coupling the first two MIs. The fundamental solution $W$ for this block, i.e., the solution~to
\begin{equation}
 \partial_{m_t^2}W=B\cdot W \,,
\end{equation}
can be found by converting the $2\times 2$ system of first order DEs into a single second-order DE
\begin{equation} \label{eq:periodDE}
 L_{m_t^2}^{(2)}\psi_i\equiv\left[\partial_{m_t^2}^2+a_1(\vec{x})\partial_{m_t^2}+a_0(\vec{x})\right]\psi_i=0 \,,
\end{equation}
by standard methods. The solutions to this equation are the two periods in \cref{eq:periods}. We then can convert back to the $2\times 2$ system to write $W$ in terms of the periods.

Following Ref.~\cite{Gorges:2023zgv}, we then split $W$ into a lower-triangular semi-simple part and an upper-triangular unipotent part, i.e.~$W=W^{\text{ss}} \cdot W^{\text{u}}$, with
\begin{equation} \label{eq:Wss}
W^{\text{ss}}=\begin{pmatrix}
               \psi_1 & 0 \\
               \psi_1' & \frac{\det(W)}{\psi_1}
              \end{pmatrix}\quad\text{and}\quad W^{\text{u}}=\begin{pmatrix}
               1 & \frac{\psi_2}{\psi_1} \\
               0 & 1
              \end{pmatrix}\,,
\end{equation}
and rotate the first two MIs by the inverse of the semi-simple part. 
Although $W$ depends on both $\psi_1$ and $\psi_2$ (and their derivatives), its semi-simple part can in general be written in terms of a single solution, say $\psi_1$ (and its $m_t^2$-derivative, $\psi_1'$), thanks to the fact that $\det(W)=\psi_1\psi_2'-\psi_2\psi_1' $
is a rational function of $\vec{x}$ due to the Legendre relation for elliptic integrals.
The freedom of choosing which solution of $L_{m_t^2}^{(2)}$ appears in $W^{\text{ss}}$ and thus in the canonical DEs is unphysical and the result for the original basis $\vec{\mathcal{I}}$ does not depend on it.

After this basis change and a rescaling by suitable factors of $\eps$, the homogeneous DEs in $m_t^2$ are now linear in $\eps$, with the leading term being lower triangular,
\begin{equation}\label{eq:linear-A-ell}
 A^{\text{ss}}_{m_t^2,\text{hom.,ell.}}=\begin{pmatrix}
  0 & 0 & 0 \\
  \star & 0 & 0 \\
  \star & \star & 0
 \end{pmatrix} + \eps \begin{pmatrix}
  \star & \star & \star \\
  \star & \star & \star \\
  \star & \star & \star
 \end{pmatrix} \,.
\end{equation}
Achieving $\eps$-factorization therefore amounts to integrating out the remaining elements at $\eps=0$~\cite{Gehrmann:2014bfa}. 
However, as was the case for several examples in Ref.~\cite{Gorges:2023zgv}, not all elements are total derivatives in $m_t^2$.
As a result, we need to introduce a new function, $G_1$, defined through its $m_t^2$-derivative, as $\partial_{m_t^2} G_1 = \bigl(A^{\text{ss},\eps=0}_{m_t^2,\text{hom.,ell.}}\bigr)_{2,1}$, which is linear in $\psi_1$ and $\psi_1'$ (the same as in $W^{\text{ss}}$), and rational in $\vec{x}$.

Having worked out the intricate dependence of the resulting transformation on all kinematic variables (see the Appendix), we can successively check the $\eps$-factorization of the homogeneous DEs not only in $m_t^2$, but also in the other variables.
We find that, thanks to our careful choice for the solutions of the appearing differential operators, these are automatically $\eps$-factorized, so that no further ($m_t^2$-independent) transformations are required.

To achieve the same for the inhomogeneous blocks of the connection matrix, we again integrate out the elements at $\eps=0$, introducing three additional functions, dubbed $G_2$, $G_{3,\pm}$. Their partial derivatives are linear in $\psi_1$ and $\psi_1'$, but now additionally contain various square roots: they are odd w.r.t.~$\sqrt{\Delta_{32}}$, and $G_{3,\pm}$ originate from the coupling of the two sectors in \cref{fig:probSec} to each other, and contain $\sqrt{\Delta_5}$ as well as $\sqrt{n_{\pm}}$, respectively.

Similarly to the first step, these transformations suffice to $\eps$-factorize the DEs in all variables, even though they were derived by only considering the derivative w.r.t.~$m_t^2$.
Moreover, while from the point of view of the DEs the new functions $G_i$ are defined through their partial derivatives, we were also able to express them explicitly as complete elliptic integrals of the third kind by considering different Baikov representations~\cite{Baikov:1996iu,Frellesvig:2017aai} and performing the integrals appearing in them, see the Appendix.
This result will be useful in view of the numerical evaluation of the solution, which we leave to future work.

\section{Analytic Structure}
\label{sec:AnalyticStructure}

The DEs for the basis $\vec{\cal J}$ constructed in the previous section take the form
\begin{align} \label{eq:canonicalDEs}
 \dd \vec{\cal J}(\vec{x}, \eps) = \eps \, \sum_{i=1}^{112} M_i \, \omega_i(\vec{x}) \cdot  \vec{\cal J}(\vec{x}, \eps)\,,
\end{align}
where $M_i$ are constant rational matrices, and $\omega_i(\vec{x})$ are $\mathbb{Q}$-linearly independent differential one-forms.
We verified the linear independence either by treating the elliptic functions ($\psi_1$, $G_i$) as independent variables---i.e., we view the entries of the connection matrix as rational functions of $\vec{x}$, $\psi_1$ and $G_i$---or by evaluating them numerically.
We expressed the 83 $\omega_i(\vec{x})$'s free of elliptic functions as $\dd \log W$ for some algebraic function $W$.
This required constructing 13 new $\dd \log$'s involving the nested square roots, besides the ones of Ref.~\cite{Badger:2024fgb}.
We chose the 29 one-forms containing elliptic functions from the entries of the connection matrix, and wrote them in terms of differentials of $\vec{x}$.
The definition of the MIs $\vec{\cal J}$ contains overall (nested) square roots and transcendental functions.
This induces two gradings on the one-forms.

First, it is well known that one-forms and MIs can be chosen to be either even or odd upon flipping the sign of each square root.
While this holds also for $\sqrt{n_{\pm}}$, flipping the interior root---$\sqrt{\Delta_5}$---exchanges $\sqrt{n_{+}}$ and $\sqrt{n_{-}}$, leading to a more complicated behavior.
One way to generalize the even/odd parity to a picture that includes nested square roots is to view it as the trivial/sign representation of $\mathbb{Z}_2$, the group formed by the operations of flipping each square root.
The relevant group for the nested square roots $\sqrt{n_{\pm}}$ is $\mathbb{Z}_2 \times \mathbb{Z}_2$: one $\mathbb{Z}_2$ flips the sign of the exterior root, the other of the interior one.\footnote{The mathematical framework is that of Galois theory. The Galois group of a polynomial is, roughly speaking, the group of unique permutations of its roots. Square roots are roots of quadratic polynomials whose Galois group is $\mathbb{Z}_2$.
The nested square roots $\pm \sqrt{n_{\pm}}$ are instead roots of $P(z) = z^4 + a z^2 + b$ (for $a$, $b$ functions of $\vec{x}$), whose Galois group is $D_4 \simeq \mathbb{Z}_2 \times \mathbb{Z}_2$.
See also refs.~\cite{Pogel:2024sdi,Duhr:2024xsy} for similar applications of Galois theory to Feynman integrals.}
For the exterior root we have the usual even/odd representation.
For the interior root, instead, we have a two-dimensional representation of $\mathbb{Z}_2$ generated by the Pauli matrix $\sigma_1$, acting on \emph{duplets}.
In other words, we arrange all objects into duplets whose elements transform into each other upon flipping $\sqrt{\Delta_5}$.
Examples of such duplets are $(\sqrt{n_{+}}, \sqrt{n_{-}})$, $(\mathcal{J}_{19},\mathcal{J}_{20})$, $(G_{3,+}, G_{3,-})$, and $(\omega_4, \omega_5)$.
This duplet structure constrains the form of the DEs, as
\begin{align}
 M_a = \Sigma \cdot M_b \cdot \Sigma^{-1}\,,
\end{align}
for every duplet $(\omega_a, \omega_b)$, where $\Sigma$ is the matrix representing the $\sqrt{\Delta}_5$-flip on the MIs.

The grading of one-forms also informs the construction of a $\dd \log$ representation, when that is possible.
While for $\dd \log$'s containing square roots there is a well understood procedure~\cite{Heller:2019gkq,Zoia:2021zmb}, results involving nested square roots are scarce, and it is in general unclear whether a $\dd \log$ representation is possible~\cite{FebresCordero:2023pww}. 
We succeeded for all one-forms free of elliptic functions, by generalising the methods applied to square roots.
For example, for a $\sqrt{\Delta_5}$-duplet of one-forms $\omega_{\pm}$ odd w.r.t.\ $\sqrt{n_{\pm}}$ and one of the square roots, say $\sqrt{\beta}$, we make the ansatz
\begin{align} \label{eq:letter_ansatz}
\omega_{\pm} = \dd \log \left(\frac{c_1 \pm c_2 \sqrt{\Delta_5} + c_3 \sqrt{\beta} \, \sqrt{n_{\pm}}}{c_1 \pm c_2 \sqrt{\Delta_5} - c_3 \sqrt{\beta} \, \sqrt{n_{\pm}}} \right) \,,
\end{align}
which manifests the grading.
We then constrain the polynomials $c_i$ by imposing that the singular locus is that of the target one-form.
This amounts to solving quartic equations, for which we adopt ad-hoc methods leveraging the sparsity of the system and of the solution.
In some cases, it is easier to integrate the one-form in one variable, setting all others to numbers and rationalising one of the roots, and use the result to constrain the ansatz.
Interestingly, the polynomial $C \coloneqq n_+ n_-$ does not appear as the argument of a $\dd \log$.
This observation is relevant in view of the program aimed at determining the logarithmic one-forms from the Landau equations~\cite{Bjorken:1959fd,Landau:1959fi,10.1143/PTP.22.128,Dennen:2015bet,Prlina:2018ukf,Mizera:2021icv,Hannesdottir:2021kpd,Lippstreu:2022bib,Fevola:2023fzn,Dlapa:2023cvx, Fevola:2023kaw, Caron-Huot:2024brh, He:2024fij,Helmer:2024wax}: $C$ appears in the Landau discriminant, 
but only factoring it as $C = (a+b \sqrt{\Delta_5}) (a-b \sqrt{\Delta_5})$, for polynomials $a$ and $b$, reveals $n_{\pm} = a \pm b \sqrt{\Delta_5}$ as possible arguments of square roots.

The second grading, that we dub \emph{elliptic grading}, is associated with the elliptic functions.
By grading $+1$ the periods and thus the $G_i$'s, all MIs and one-forms gain a uniform grade ranging from $-2$ to $+2$, as already observed in other cases.
Note that a function having elliptic grade $0$ may still contain elliptic functions, as ratios with $\psi_1$ in the denominator also appear, e.g.~$G_1/\psi_1$.
We stress that the emergence of this grading is here a mere observation, that we use to classify MIs and one-forms; its implications are left for future study.
Additionally, we note that elliptic one-forms appear also in the entries of the connection matrix coupling MIs of sectors that are not elliptic on their maximal cut, and that $\psi_1'$ is absent from the DEs, although present in the transformation~$T$.

Finally, we verified that the DEs in \cref{eq:canonicalDEs} are canonical according to the local definition
given above.
They contain transcendental functions, in particular $\psi_1$.
What matters for the $\eps$-factorization is that $\psi_1$ satisfies the DE~\eqref{eq:periodDE};
any linear combination of $\psi_1$ and $\psi_2$ would work, and this freedom extends to the $G_i$'s.
However, in the neighborhood of certain singular points, one solution to \cref{eq:periodDE} is holomorphic while the other has logarithmic singularities;
which one depends on the point.
We verified that, for each singular point, there is a solution to \cref{eq:periodDE} such that all one-forms have at most simple poles.
In practice, we pull the one-forms back to paths parameterized by one variable, and expand the elliptic functions around each singular point via Frobenius' method~\cite{Frobenius,Ince}, choosing the holomorphic solution.

Our analytic results can be found at Ref.~\cite{zenodo}.

\section{Conclusion and Outlook}
\label{sec:conlusion}

We derived canonical DEs for the two-loop, five-point integral family involving elliptic functions that contributes to the leading color NNLO QCD corrections to top-pair production with a jet at hadron colliders.
Together with the boundary values of Ref.~\cite{Badger:2024fgb}, this makes it straightforward to write the solution analytically in terms of iterated integrals with closed-form kernels~\cite{Chen:1977oja}.
This result finalizes the construction of canonical DEs for all integral families needed for this prediction~\cite{Badger:2022hno, Badger:2024fgb, Badger:2024dxo} and opens several interesting research avenues.

From a theoretical standpoint, this achievement opens the door to investigating the analytic structure of amplitudes involving higher-genus geometries in a high-multiplicity regime. In particular, recent studies have uncovered striking patterns of cancellations among combinations of iterated integrals involving elliptic functions and even functions defined over Calabi-Yau varieties~\cite{Duhr:2024bzt,Forner:2024ojj,Marzucca:2025eak,Abreu:2022cco,Delto:2023kqv,Becchetti:2025rrz}, leading to remarkable simplifications at the amplitude level. 
It would be interesting to establish whether similar simplifications occur in the present context, determine their potential implications for phenomenology, and explore whether they can be linked to the elliptic grading analysed in this work.

The generalization of the even/odd parity to incorporate nested square roots and the constraints this imposes on Feynman integrals through their DEs are other important insights raised in this work. 
This aspect is closely tied to the possibility of expressing DEs containing nested square roots in $\dd \log$ form. 
Our results lay a strong foundation to extend to nested square roots the active study of symbol letters~\cite{Dennen:2015bet, Prlina:2018ukf, Drummond:2019cxm,Heller:2019gkq, Chicherin:2020umh, Mizera:2021icv, Hannesdottir:2021kpd, Zoia:2021zmb, Lippstreu:2022bib, Morales:2022csr, He:2022ctv, Bossinger:2022eiy, Yang:2022gko, He:2022tph, Dlapa:2023cvx, Fevola:2023kaw, Chen:2023kgw, Fevola:2023fzn, He:2023umf, Caron-Huot:2024brh, Jiang:2024eaj, He:2024fij, Helmer:2024wax,Aliaj:2024zgp}, which underpins many bootstrap programs (see Ref.~\cite{Hannesdottir:2024hke} for a review).

Another primary challenge is to obtain an analytic representation of the MIs that not only reveals simplifications in the amplitude but also enables efficient numerical evaluation. 
As of now, the best available approach to evaluate the MIs considered here is that of generalized series expansions~\cite{Moriello:2019yhu} applied to non-canonical DEs~\cite{Badger:2024fgb,Badger:2024dxo}; while flexible, this method suffers from slow evaluation times w.r.t.\ what is typically needed for Monte-Carlo integration.
Recent progress in expressing MIs in terms of independent special functions~\cite{Gehrmann:2018yef,Chicherin:2020oor,Chicherin:2021dyp,Abreu:2023rco,Gehrmann:2024tds} together with a specialized approach to their evaluation~\cite{Caron-Huot:2014lda,Chicherin:2020oor} have led to a surge of NNLO QCD predictions for $2 \to 3$ LHC processes with massless virtual particles~\cite{Chawdhry:2019bji,Kallweit:2020gcp,Chawdhry:2021hkp,Czakon:2021mjy,Badger:2021ohm,Chen:2022ktf,Alvarez:2023fhi,Badger:2023mgf,Hartanto:2022qhh,Hartanto:2022ypo,Buonocore:2022pqq,Catani:2022mfv,Buonocore:2023ljm,Mazzitelli:2024ura,Devoto:2024nhl,Biello:2024pgo,Buccioni:2025bkl}.
Our results set the necessary ground to study the extension of these techniques to elliptic integrals in a context where this is essential to obtain theoretical predictions.
In fact, the complexity of the Monte Carlo phase-space integration sets substantially higher efficiency demands for $2\to 3$ processes as compared to lower multiplicities.
The extension of the techniques above would thus have a transformative impact, and remove one of the main bottlenecks to improve the predictions for a wide class of processes.
More general strategies based on series expansions are also under study for iterated integrals with higher-genus geometries~\cite{Badger:2022mrb,Delto:2023kqv,Duhr:2024bzt,Forner:2024ojj,Becchetti:2025rrz}, and their potential merits further exploration.

Beyond being a critical step towards achieving NNLO QCD corrections for an important LHC process, our results prove that the methodology under development by the community has reached the maturity to tackle the computation of elliptic Feynman integrals with five-particle kinematics.
This milestone is important also in view of other massive $2\to 3$ amplitudes, which are the object of active research~\cite{Buccioni:2023okz,FebresCordero:2023pww,Agarwal:2024jyq,Becchetti:2025osw,Becchetti:2025qlu} and play a crucial role in LHC phenomenology.
Furthermore, the mathematical challenges addressed in this work are relevant in other research areas, including gravitational waves~\cite{Dlapa:2023hsl,Dlapa:2024cje,Driesse:2024feo,Bern:2024adl}, string theory~\cite{Berkovits:2022ivl} and pure mathematics, underscoring the broad impact of our findings.

\acknowledgments

We thank S.~Badger, C.~Nega, V.~Sotnikov, L.~Tancredi and F.~Wagner for useful discussions, the organizers of `\href{https://indico.dfa.unipd.it/event/847/}{MathemAmplitudes 2023},' and CERN for hospitality.
We are grateful to S.~Badger, J.~Henn and V.~Sotnikov for useful comments on the draft.
This project received funding from the European Union’s Horizon Europe research and innovation programme under the Marie
Skłodowska-Curie grant agreement No.~101105486
and ERC Starting Grant No.~101040760 \emph{FFHiggsTop},
and from the Swiss National Science Foundation under the Ambizione grant No.~215960.


\appendix

\section{Explicit expressions for the \texorpdfstring{$G_i$'s}{Gi}}
\label{sec:third-kind}

In this appendix we explain how we obtained expressions for the functions $G_i$ appearing in the canonical DEs in terms of complete elliptic integrals of the third kind. The procedure is similar to the one described in Ref.~\cite{Gorges:2023zgv} around Eq.~(3.53), with the difference that we cannot always cut all propagators since we are also interested in functions that appear in the inhomogeneous parts of the DEs.

We start with the functions $\psi_i$ and $G_1$ that appear in the elliptic sector spanned by the basis given in Eq.~(3.31) of Ref.~\cite{Badger:2024fgb}.\footnote{We find it convenient to swap the second and third integral to find the structure given in \cref{eq:elliptic-ep0,eq:elliptic-ep0bis}.} As discussed there, the maximal cut of the first integral gives rise to the two periods, i.e.
\begin{equation}\label{eq:period-cut}
 \mathrm{MaxCut}\left[I_{1,1,0,1,1,1,0,1}^{0,0,0}\right]_{\eps=0}\propto\int\frac{\dd z}{y} \,,
\end{equation}
where $z$ is the Baikov variable associated to the eighth propagator and $y^2=(z-e_1)(z-e_2)(z-e_3)(z-e_4)$ is the elliptic curve.
The roots $e_i$ are functions of $\vec{x}$ and can be found in Ref.~\cite{Badger:2024fgb}.
We therefore see why the (properly normalized) periods $\psi_1$ and $\psi_2$ defined in \cref{eq:periods} are solutions to the second-order differential operator $L_{m_t^2}^{(2)}$ annihilating the first integral at $\eps=0$.

Similarly, the transformation to canonical form introduces $G_1$ as the solution to a third-order differential operator (related to $L_{36}$ discussed in Ref.~\cite{Badger:2024fgb}). In more detail, the DEs in \cref{eq:elliptic-ep0bis} can be rewritten as a third-order differential operator annihilating the third integral, $I_{1,1,0,1,1,1,0,1}^{0,0,0}[(k_1+p_1)^2]$, where the term in the square brackets is a numerator under the integral sign. The general solution for this DE is given by
\begin{equation}
 \begin{aligned}
  I_{1,1,0,1,1,1,0,1}^{0,0,0}\left[(k_1+p_1)^2\right]_{\eps=0}&\propto c_1 \left(4m_t^2\psi_1^2+G_{1,1}\right)\\
  & \hspace{-50pt}+c_2 \left(4m_t^2\psi_2^2+G_{1,2} \right)+c_3+\ldots \,,
 \end{aligned}
\end{equation}
where $G_{1,1}$ and $G_{1,2}$ are the function $G_1$ evaluated on the same integration cycle as the period $\psi_1$ and $\psi_2$, respectively, the $c_i$ are yet undetermined boundary constants, and the ellipsis indicates terms coming from subsectors.
On the other hand, the maximal cut of this integral is
\begin{equation}
 \mathrm{MaxCut}\left[I_{1,1,0,1,1,1,0,1}^{0,0,0}\left[(k_1+p_1)^2\right]\right]_{\eps=0}\propto\int\frac{\dd z(z-e_1)}{y} \,,
\end{equation}
that evaluates to a complete elliptic integral of the third kind $\Pi$ with argument either $\kappa^2$ or $1-\kappa^2$ (defined below \cref{eq:K}), depending on the integration cycle.
For example, we have that
\begin{equation}
 \int_{e_2}^{e_3}\frac{\dd z(z-e_1)}{y}=2(e_1-e_2)\frac{\mathrm{\Pi}\left(\frac{e_3-e_2}{e_3-e_1},\kappa^2\right)}{\sqrt{(e_3-e_1)(e_4-e_2)}} \,,
\end{equation}
where
\begin{equation}
 \Pi\left(a,\kappa^2\right)=\int_0^1\frac{\dd t}{(1-at^2)\sqrt{(1-t^2)(1-\kappa^2t^2)}} \,.
\end{equation}
This immediately relates $G_1$ to either $\mathrm{\Pi}(\frac{e_3-e_2}{e_3-e_1},\kappa^2)$ or $\mathrm{\Pi}(\frac{e_1-e_2}{e_1-e_3},1-\kappa^2)$ depending on which period is chosen in the semi-simple part $W^{\rm ss}$ in \cref{eq:Wss}.

The remaining three functions, $G_2$ and $G_{3,\pm}$ are instead related to the inhomogeneous parts of the DEs.
For example, $G_2$ appears in the general solution of a third-order differential operator annihilating the following integral that couples to the elliptic sector:
\begin{equation}\label{eq:G2-integral}
 \begin{aligned}
  I_{1,1,0,1,1,1,1,1}^{0,0,0}\left[(k_2+p_2)^2\right]_{\eps=0}&\propto c_1\\
  &\hspace{-40pt}+c_2\left(8 \, G_{2,1}+4 \, m_t^2 \, \frac{\psi_1}{\Delta_{32}}\right)\\
  &\hspace{-40pt}+c_3\left(8 \, G_{2,2}+4 \,m_t^2 \, \frac{\psi_2}{\Delta_{32}}\right)+\ldots \,,
 \end{aligned}
\end{equation}
where the ellipsis indicates terms coming from subsectors of the elliptic sector.
As before, $G_{2,i}$ is the function $G_2$ evaluated on the same integration cycle as $\psi_i$.
Note that, on the maximal cut of the integral (i.e., additionally cutting the seventh propagator), the elliptic functions in \cref{eq:G2-integral} disappear and the solution is simply $c_1$.
As in the previous case, we proceed by comparing \cref{eq:G2-integral} to the Baikov representation:
\begin{equation}\label{eq:G2-Baikov}
 \begin{aligned}
  &\mathrm{Cut}_{\text{ell.}}\left[I_{1,1,0,1,1,1,1,1}^{0,0,0}\left[(k_2+p_2)^2\right]\right]_{\eps=0}\\
  &\hspace{50pt}\propto\int\frac{\dd z_7\dd z_9\sqrt{m_t^2+z_9}}{z_7\sqrt{f_1(z_7,z_9)}\sqrt{f_2(z_7,z_9)}} \,,
 \end{aligned}
\end{equation}
where $\mathrm{Cut}_{\text{ell.}}$ denotes the cut of the elliptic sector, $f_1(z_7,z_9)$ is linear in $z_7$ and $z_9$, and $f_2(z_7,z_9)$ is quadratic in $z_7$ and $z_9$. After a change of variables, $\{z_7,z_9\}\to\{x,y\}$, that rationalizes both square roots in the denominator, the integral in $x$ can be performed to show that the integrand of \cref{eq:G2-Baikov} has the form
\begin{equation}
 \dd\log[\alpha_1(x,y)]\,\dd y\, g_1(y)+\dd\log[\alpha_2(x,y)]\,\dd y\, g_2(y) \,.
\end{equation}
Similarly to the procedure described in Appendix~A of Ref.~\cite{Badger:2024fgb}, we now  analyse $g_1(y)$ and $g_2(y)$, and ignore the $\dd \log$'s.
We find that $g_2(y)$ can also be written in $\dd \log$ form with unit coefficient.
This term corresponds to the constant $c_1$ in \cref{eq:G2-integral}, and indeed disappears when subtracting the residue at $z_7=0$ before integration.
On the other hand, integrating the function $g_1(y)$ gives rise to one of two complete elliptic integrals of the third kind, depending on the integration cycle. Comparing to \cref{eq:G2-integral} relates these to the functions $G_{2,1}$ and $G_{2,2}$.

In practice, rationalising the appearing square roots and integrating in the Baikov variables can become computationally expensive.
Therefore, we often perform these computations while substituting generic rational numbers for some of the kinematic variables.
The algebraic functions that appear as coefficients or arguments of the elliptic integral of the third kind can then often be guessed by comparing them to factors that already appear in the non-canonical DEs.
For example, we observe that expressions of the form $a+b\sqrt{f}$ typically factor in terms of denominator factors of the DEs when multiplied with their conjugate:
\begin{equation}\label{eq:square-root-factor}
 (a+b\sqrt{f})(a-b\sqrt{f})=a^2-b^2f \,.
\end{equation}
This makes it straightforward to extract the analytic expression for the l.h.s.\ of~\eqref{eq:square-root-factor} once the r.h.s.\ is identified.

In addition, we find that all terms integrating to complete elliptic integrals of the third kind satisfy a DE of the form
\begin{equation} \label{eq:PiPrime}
 \begin{aligned}
  \left(\Pi(a,k)\frac{\sqrt{a-1}\sqrt{a-k}}{\sqrt{a}}\right)'&=\mathrm{K}(k)\frac{\sqrt{a-k}a'}{2\sqrt{a-1}a^{3/2}}\\
  &\hspace{-70pt}+\mathrm{E}(k)\frac{(a-1)k'-(k-1)a'}{2\sqrt{a-1}\sqrt{a}\sqrt{a-k}(k-1)} \,,
 \end{aligned}
\end{equation}
where the prime indicates a partial derivative w.r.t.\ some kinematic variable.
We emphasize that the r.h.s.\ of \cref{eq:PiPrime} does not contain elliptic integrals of the third kind.
For example, this relation can be used to determine the prefactor of $\Pi(a,k)$ when the arguments $a$ and $k$ have been identified.
Moreover, once the result for one integration cycle is known, the replacement
\begin{equation}
 \begin{aligned}
  \mathrm{K}(k)&\to\mathrm{K}(1-k),\\
  \Pi(a,k)&\to\frac{a}{a-1}\Pi(1-a,1-k)-\frac{1}{a-1}\mathrm{K}(1-k) \,,
 \end{aligned}
\end{equation}
is useful for identifying the functions for the other integration cycle, since it preserves the partial derivatives of the $G_i$ when written in terms of the periods.

\bibliography{bibliography}

\end{document}